\documentstyle[aps,preprint,eqsecnum,epsf,tighten]{revtex}
\newcommand{\APSREF}{1}

\newcommand{\dal}{\partial^{\mu}\partial_{\mu}}
\newcommand{\Dal}{{\,\lower 1.0pt\vbox{\hrule \hbox{\vrule height 0.3cm
\hskip 0.28 cm \vrule height 0.3 cm}\hrule}\,}}

\ifnum\APSREF=1

\newcommand{\physrevd}[3]{Phys. Rev. D{\bf #1}, #2 (#3)}
\newcommand{\physrep}[3]{Phys. Rep. {\bf #1}, #2 (#3)}

\newcommand{\npb}[3]{Nucl. Phys. B{\bf #1}, #2 (#3)}
\newcommand{\physrevlett}[3]{Phys. Rev. Lett. {\bf #1}, #2 (#3)}

\else

\newcommand{\physrevd}[3]{Phys. Rev. D{\bf #1} (#3) #2 }
\newcommand{\physrep}[3]{Phys. Rep. {\bf #1} (#3) #2 }

\newcommand{\npb}[3]{Nucl. Phys. B{\bf #1} (#3) #2 }
\newcommand{\physrevlett}[3]{Phys. Rev. Lett. {\bf #1} (#3) #2 }

\fi

\newcommand{\splash}[1]{{#1\mkern-9.0mu /}}          
\def\'#1{\if#1i{\accent 19 \i}\else{\accent 19 #1}\fi}

\def\dal{\,\lower0.9pt\vbox{\hrule \hbox{\vrule height 0.25 cm \hskip 0.25 cm 
\vrule height 0.25 cm}\hrule}\,}
\def\dalo{\,\lower0.9pt\vbox{\hrule \hbox{\vrule height 0.2 cm \hskip 0.2 cm 
\vrule height 0.2 cm}\hrule}\,}
\def\'#1{\if#1i{\accent 19 \i}\else{\accent 19 #1}\fi}

\begin{document}

\title{Neutrino damping rate at finite temperature and density}

\author{Eduardo  S. Tututi\footnote{Email: tututi{@}zeus.ccu.umich.mx}}

\address{Escuela de Ciencias F\'isico Matem\'aticas,  Apartado  Postal 2-71, 58040\\
Universidad Michoacana,  Morelia Mich., M\'{e}xico}

\author{Manuel  Torres\footnote{Email: torres@fisica.unam.mx}}

\address{Instituto de F\'isica,  Apartado  Postal 20-364, 01000,\\Universidad Nacional Aut\'onoma de  M\'exico, D.F., M\'exico}

\author{Juan Carlos  D'Olivo\footnote{Email: dolivo@nuclecu.unam.mx}}

\address{Instituto de Ciencias Nucleares, Apartado  Postal 70-534, 04510,\\  Universidad Nacional Aut\'onoma de  M\'exico M\'exico, D.F., M\'exico}


\maketitle

\begin{abstract}
A first principle derivation is given of the neutrino damping rate in real-time
thermal field theory. Starting from the discontinuity of the neutrino self 
energy at the two loop level, the damping rate can be expressed as integrals
over space phase of amplitudes squared, weighted with statistical factors that 
account for the possibility of particle absorption or emission from the medium.
Specific results for a background composed of neutrinos, leptons, protons and neutrons
are given. Additionally,  for the real part of the dispersion relation we
discuss the relation between the results obtained from the thermal field 
theory, and those obtained by the thermal average of the forward scattering 
amplitude.
\end{abstract}

\draft

\section{Introduction}\label{introduction}
\indent

A neutrino propagating in matter do not longer respect the vacuum energy momentum relation.
The modification of the neutrino dispersion relation  is caused by their coherent interaction with the particles  in the background and can be accounted in terms of an index of refraction \cite{lang}  or an effective 
potential  \cite{bet}. The topic of neutrino propagation in matter became of prime relevance
after Wolfenstein \cite{wolf} study of the  neutrino refractive index in matter,  and later when 
Mikheyev-Smirnov \cite{ms} recognized the resonant neutrino flavor oscillations triggered by matter effects.
The Mikheyev-Smirnov  effect has become the most popular explanation of the solar neutrino deficit   \cite{bah1}.
 
In general, the neutrino dispersion relation  is a complex
function $\omega_\kappa = \omega(\vec{\kappa})$.
According to the Thermal Field Theory (TFT)  matter contributions  to the
real and imaginary parts of  $\omega(\vec{\kappa})$ arise from the temperature and density-dependent  part of  the  neutrino self-energy. 
To  leading order in $(g^2/M_W^{2})$ the real part  of the dispersion relation is
proportional to the particle-antiparticle  asymmetry in the background.
If the asymmetry is small or zero the imaginary part of $\omega(\kappa)$ and corrections of order $g^2/M_W^{4}$ to the real part may be important because they do not depend on the differences between the number of particles and antiparticles. This may be the case in the early Universe, 
when the  medium was probably nearly  CP-symmetric.

Special attention has been given to the  calculations of the $O(g^2/M_W^{4})$ 
corrections  to the real part of the neutrino  dispersion relations  \cite{notz,enqv,dnt}.
Within the framework of the TFT these corrections  arise  from the  momentum-dependent terms of the
boson propagator in the self-energy diagrams. 

The imaginary  part of the index of refraction  for neutrinos propagating in a CP-symmetric plasma  
composed of  electrons, neutrinos and their  anti-particles has been  considered in references 
\cite{notz,enqv2,lang2}. These calculations have been based upon the computation of  the neutrino reaction rates, assuming massless background fermions. 
In our opinion  the relation of the neutrino damping rate to the  self-energy discontinuties 
deserves further consideration. Additionally this work  addresses the effects of the nucleons background contributions and the fermion mass correction. 

A systematic method to compute the damping rate from the imaginary part of
the self energy can be formulated in terms of  the Cuttosky thermal rules.
 Weldon \cite{wel1} and Kobes \cite{kob1}
proved that the imaginary part of the self energy can be organized in a form that includes square 
of amplitudes of the various processes obtained from the cuts of the self energy and 
weighted with the appropriated statistical factors.  The examples discussed in those papers 
are always at the one loop level. As it shall be discussed, the calculation of the 
neutrino damping rate require to consider the self energy at the two loop level,
the interpretation in terms of square  of amplitudes of the allowed  processes will be proved to  remain  valid.  The approximations required to recover the results obtained from the optical theorem will be clearly
stated.

In this work we use the method of real-time thermal field theory  to carry out
a carefully calculation of the imaginary part of the neutrino dispersion relation
in a medium composed of electrons, protons,  neutrons,
neutrinos and their  anti-particles. As already mentioned the
contributions to the imaginary part of the neutrino self-energy vanishes 
at the  one loop level,  so we  have to consider the contributions at the two loops
level.

The paper is organized as follows. In the next section
we briefly review those ingredients of the FTF formalism that are required
to accomplish our calculations. In  Section  \ref{real part} we re-derive the real part of the 
dispersion relation of a massless neutrino that propagate in a thermal 
background. We prove that utilizing  the methods of finite -temperature field theory (at the $g^2$ order),    the neutrino  effective potential reduces to the   thermal average of neutrino forward scattering amplitude.
The calculation of the imaginary part of the dispersion relation 
is presented in Section  \ref{imaginary part}. 
The  neutrino damping rate is extracted from the discontinuity of self-energy
at the two loop level, it is   expressed in terms of  integrals
over space phase of amplitudes squared, weighted with statistical factors that 
account for the possibility of particle absorption or emission from the medium.
Specific results for a background composed of neutrinos, leptons, protons and neutrons
are given. The last section contains a summary of
our main results.

\section{Basic formalism}\label{basic formalism}

The relevant quantity is the self-energy $\Sigma$, which embodies the 
effects of the background on a  neutrino that propagates through it. 
According  to the real-time formalism of the TFT \cite{lands,lebellac,nieves},
the real and imaginary part  of  $\Sigma$ are given by the formulas
\begin{equation}\label{ein1}
{\rm Re} \, \Sigma = {\rm Re}\, {\Sigma}_{11}\,, 
\end{equation}
\begin{equation}\label{ein2}
{\rm Im} \, \Sigma = {{\Sigma}_{12}\over{2i\left(\eta_f(k\cdot u)-
\theta(-k\cdot u)\right)}} \,,
\end{equation}
where  $\Sigma_{ab}$  (a,b = 1,2) are the complex elements of the
 $2\times2$ self-energy  matrix  to be computed utilizing the Feynman 
rules of the theory. $\theta$ is the step function 
and 
\begin{equation}\label{ein3}
\eta_{f}(k\cdot u)=\left[ \theta(k\cdot u)   n_f (x_k) \, + \,
 \theta(-k\cdot u)  n_f (-x_k)  \right]  
\, ,
\end{equation}
where the thermal distribution is given by 
\begin{equation}\label{ein4}
 n_f (x_k) = {1 \over { e^{x_k} + 1} }\,, 
\end{equation}
with $ x_k= \beta (k \cdot u - \mu_f)$.
Here, $\beta$ is the inverse of the temperature and
$\mu_f$ is the chemical potential
associated with the fermion $f$. 
We have introduced the  velocity  four-vector of the background $u^\mu$.
In  its own rest frame  $u^\mu =  (1,{\bf 0})$ and  the components of the neutrino 
momentum $k^\mu$ are $k^\mu  = (\omega, \vec{\kappa})$.

In the presence of a  medium the chiral nature of the 
neutrino interactions implies that the  self-energy of a massless (left-handed) neutrino is of the
form  \cite{wel2}
\begin{equation}\label{ein5}
\Sigma (k)=(a\splash{k}+ b\splash{u} )L
\, ,
\end{equation}
where $L= (1 - \gamma_5)/2$ and  $a$, $b$  are complex functions of the scalars

\begin{equation}\label{ein6}
\omega=k\cdot u,  \ \ \ \ \ \ \kappa = \sqrt{\omega^2-k^2}\,.
\end{equation}

 In this case, 
the Dirac equation for the spinor $U$  of the neutrino mode in the medium is
\begin{equation}
\label{ein7}
\left(\splash{k}-\Sigma \right)U=0\,,
\end{equation}
which has non-trivial 
solutions  only for those values of $\omega$ and $\kappa$ such that $V^2=0$,
with $V_{\mu}=(1-a)k_{\mu}-b u_{\mu}$. 
Then, the dispersion relations $\omega_\kappa$ of the 
neutrino and antineutrino modes  are obtained  as the solutions of 
\begin{equation}
\label{ein8}
f(\omega_\kappa, \kappa) = 0,
\end{equation}
and
\begin{equation}
\label{ein9}
\overline f(\omega_\kappa, \kappa) = 0\,,
\end{equation}
where 
\begin{eqnarray}\label{ein10}
f(\omega, \kappa) &=& (1 - a)(\omega - \kappa) - b\,,\nonumber\\
\overline f(\omega, \kappa)&=& (1 - a)(\omega + \kappa) - b\,.
\end{eqnarray}

In general, the solutions $\omega_\kappa$ are complex, 
as usual we write 
\begin{equation}
\label{ein11}
\omega_\kappa = \omega_r  - i \gamma/2,
\end{equation}
where both $\omega_r =  {\rm Re} \, \omega_\kappa$ and 
$\gamma= -2 \, {\rm Im}\, \omega_\kappa$ are real functions
of $\kappa$.  
A consistent
interpretation in terms of the dispersion relation for a mode propagating
through a medium is possible only if the imaginary part of
$\omega _\kappa$ is small compared to its real part.
In this case the mode can be visualized as a particle (or antiparticle)  with an energy
$\omega_r$ and a damping $\gamma$. 
Under such assumptions, each one of
Eqs.~(\ref{ein8}) and (\ref{ein9})
yields two distinct solutions, one
with positive energy and the other with negative energy, whose
physical interpretation has been discussed in detail in  Ref \cite{wel2}.
Here we will concentrate on the solution of  Eq. (\ref{ein8}) having
a positive real part, which corresponds to the neutrino
mode with energy $\omega_r$, but
similar considerations and results apply to the other 
solutions.

It is convenient to make the decompositions $a = a_r + ia_i$ and $b = b_r + ib_i$.
Then, using Eqs. (\ref{ein10}) and  (\ref{ein11}), expanding in powers of $\gamma$ and retaining only terms
that are at most linear in $\gamma, a_i$ and $b_i$, from Eq. (\ref{ein8}) we obtain 
\begin{equation}
\label{ein12}
f_r(\omega_r, \kappa) = 0 \, , 
\end{equation}
and
\begin{equation}
\label{ein13}
\frac{\gamma}{2} = \left[\frac{f_i(\omega, \kappa)}
{\frac{\partial f_r}
{\partial\omega}}\right]_{\omega=\omega_r}\,,
\end{equation}
with
\begin{eqnarray}\label{ein14}
f_r & = & (1 - a_r)(\omega - \kappa) - b_r\,,\nonumber\\
f_i & = & -a_i(\omega - \kappa) - b_i.
\end{eqnarray}
Only  approximate analytical solutions of Eq. (\ref{ein12}) are known \cite{wel2}. 
At  the one-loop level both $b_r(\omega_r, \kappa)$ and $a_r(\omega_r, \kappa)$ are of order $g^2/M_W^2$.  Therefore, to this order the energy of a massless neutrino is  
\begin{eqnarray}\label{ein15}
 \omega_r \cong \kappa +  b_r(\omega_r, \kappa){\big \vert}_{\omega_r=\kappa}\,.
\end{eqnarray}
On the other hand, as  we will show, the imaginary part of  $\Sigma$ 
vanish  to $O(g^2)$ and  to this order there are not contributions to ${\rm Im}\, \omega_\kappa$. For the perturbative solution of Eq. (\ref {ein13})  around $\omega_r = \kappa$, we have
\begin{equation}\label{ein16}
{\gamma \over 2}  \cong  -  b_i(\omega_r, \kappa){\big \vert}_{\omega_r=\kappa}
\, ,
\end{equation}
with $b_i(\omega_r, \kappa)$ of $O(g^4/M_W^4)$.
 
The matter effects on the neutrino oscillations are conveniently 
incorporated  by means of the effective potential $V$. This can be introduced
by subtracting the (vacuum) kinetic energy from the real part of the 
neutrino dispersion relation\cite{dnt}:
\begin{equation}
\label{ein17}
V \equiv \omega_r - \kappa \cong b_r(\omega_r, \kappa){\big \vert}_{\omega_r=\kappa}
\,.
\end{equation}
In the literature it is also  customary to use a refraction index, which is defined by
\begin{equation}\label{ein18}
n (\kappa) \equiv{ \kappa \over \omega_\kappa}. 
\end{equation}
In the approximation we are working on, and utilizing Eqs.  (\ref{ein11}-\ref{ein17}),  it  follows  that its real and imaginary parts are related to the effective potential and the damping rate by 
\begin{eqnarray}\label{ein19}
{\rm Re} \, n (\kappa) & \cong & 1 - {V \over \kappa}\,,\nonumber\\
{\rm Im} \, n (\kappa)& \cong &   {\gamma \over 2 \kappa}\,,
\end{eqnarray}
with $V$ and $\gamma$ given by Equations (\ref{ein17}) and (\ref{ein16}), respectively.

\section{Effective Potential}\label{real part}

According to Eq. (\ref {ein17}), for a perturbative solutions of the 
dispersion relation around the vacuum value $\omega_r = \kappa$, 
the effective potential is given by the real part of the coefficient of $\splash{u}$
in the neutrino self-energy. This is only true in the lowest order,  in general,
$V$ will also  receive contributions from $a_r$.  From Eq. (\ref{ein5}), it is 
easy to see that
\begin{eqnarray}\label{ep1}
b_r(\omega, \kappa) = {1\over4}{\rm Tr}\left\{{\splash{\lambda}{\rm Re}\Sigma}
\right\}\, .
\end{eqnarray}
with
\begin{eqnarray}\label{ep2}
\splash{\lambda} ={1\over \kappa^2} (\omega \splash{k} -  k^2 \splash{u})
\, .
\end{eqnarray}

In  Ref. \cite{dnt} the real part of the neutrino self-energy was calculated in a general gauge up to terms of order $g^2/M_W^4$. It 
was shown, that although the self-energy depends on the gauge parameter, the 
dispersion relation is independent of it.  Taking this result into account, 
for simplicity we will work in the unitary gauge. Furthermore,  for physical situations where
 the temperature is much lower than
the masses of the gauge bosons,  the propagators for the $W$ and $Z$
can be taken the same as in the vacuum, namely,  
\begin{eqnarray}\label{ep3}
 \Delta^{\alpha \beta}_B= {1\over{k^{2} - M^{2}_B}+i\epsilon}  
\left( g^{\alpha \beta} - {k^{\alpha} k^{\beta} \over { M^{2}_B}} \right)
\, ,
\end{eqnarray}
with $B=W,Z$.

 We shall assume  that neutrinos are massless,  so at the one-loop level the only
contributions to  $\Sigma$ arise from the diagrams depicted
in Fig. 1.

For a  neutrino $\nu_l$ ($l =e, \mu, \tau$) that propagates through a medium with 
a momentum  $k$, we split the different contributions to $b^l$  according to the processes in Fig. 1  as follows:

\begin{eqnarray}
\label{ep4}
b^l_r = b^l_{tad}+  b^l_Z +  b^l_W  
\, ,
\end{eqnarray}
 corresponding to the tadpole, Z-exchange, and W-exchange  contributions. In this case, the background dependent parts of each term in 
the right  hand side of the previous equation  can be worked out as 
\begin{eqnarray}\label{ep5}
b^l_{tad} &=& {1\over 4} {\rm Tr} \left\{ \splash {\lambda}  \Gamma_{\nu_l\alpha}^Z 
\right\} \Delta^{\alpha \beta}_{Z} (0) 
\sum_f\int{{{d^{4}p}\over{(2\pi)^{3}}}   {\rm Tr}  \left\{ 
( \splash{p} + m_f)\Gamma_{f\beta}^Z \right\}
\delta(p^2 - m^2_f)\, \eta_{f} (p \cdot u) }\, , \nonumber\\  
b^l_Z&=& - {1\over 4}\int{{{d^{4}p}\over{(2\pi)^{3}}} 
  {\rm Tr } \left\{ \splash{\lambda}\,
\Gamma_{\nu_{{}_l}\alpha}^{Z} \, \splash{p}\,\Gamma_{\nu_{{}_l}\beta}^Z \right\}
 \Delta^{\alpha \beta}_{Z}(k - p) \delta(p^2 )\, \eta_{\nu_l} 
(p \cdot u)} \, , \nonumber\\
b^l_W&=& - {1\over 4}\int{{{d^{4}p}\over{(2\pi)^{3}}} 
  {\rm Tr } \left\{ \splash{\lambda}
\Gamma_{\alpha}^{W} ( \splash{p} + m_l)\Gamma_{\beta}^W \right\}
 \Delta^{\alpha \beta}_{W}(k - p) \delta(p^2 - m^2_l)\, \eta_{l} (p \cdot u)} \,.
\end{eqnarray}
In  $b_W$ the charged lepton in the internal line
has the same flavor  that $\nu_l$ , while in the tadpole contribution $b_{tad}$ the summation is over all the fermion species  present in the thermal background.  In the previous expressions the vertices are given by 
\begin{eqnarray}\label{ep6}
\Gamma_{\alpha}^W & = & i {g \over {2\sqrt{2}}} \gamma_{\alpha}(1 - \gamma_5)\,,\nonumber\\  
\Gamma_{f\alpha}^Z  & = & i {g \over {2 \cos\theta_W}}\gamma_{\alpha}(X_f+Y_f\gamma_5)\,.
\end{eqnarray}
With the vector $X_l$ and axial couplings $Y_l$  given for the charged leptons by 
\begin{eqnarray}\label{ep7}
X_l & = &   -{1\over 2} +  2\sin^2\theta_W, \nonumber\\  
Y_l & = &  {1\over 2}\,;
\end{eqnarray}
for the neutrinos by 
\begin{equation}\label{ep8}
X_{\nu_l}  = - Y_{\nu_l}  =   {1\over 2} \,;  
\end{equation}
and for the nucleons by 
\begin{eqnarray}\label{ep9}
X_p & = &  {1\over 2} - 2\sin^2\theta_W, \nonumber\\  
Y_p & = & X_n = - Y_n = - {1\over 2}\,.
\end{eqnarray}

\noindent
According to Eq. (\ref{ein17})  $b^l_r$ has to be evaluated at  $\omega=\kappa$, in this case  the quantity 
$\kappa \splash{\lambda}$ in (\ref{ep2}) reduces to the energy projector for a 
massless particle in the vacuum $\splash{k}$, and may be replaced by its 
usual expression in terms of the free spinors. In general  for the external lines the spinors to be used
are the solutions of the Dirac equation in the medium  \cite{donie}. However,
within the approximation we are using, they can be approximated by the 
vacuum solutions.
For the internal fermionic lines  we substitute $ \splash{p} + m_f$ by its
corresponding energy projector.   It is now straightforward to show that any of the $b^l$ in   Eq. (\ref{ep5}) can be written in the   the form 
\begin{eqnarray}\label{ep10} 
 b^l_r (\omega,\kappa){\big \vert}_{\omega=\kappa} 
= {1 \over 2 \pi \kappa }  \langle  {\cal{M}}^l_f\,  \rangle
\,,
\end{eqnarray}
where ${\cal{M}}^l_f$ is the tree-level invariant amplitude  for the  forward 
scattering $\nu_l  f  \rightarrow \nu_l f $. The corresponding  Feynman diagrams  
are shown in Fig. 2.  Notice that these diagrams are obtained from the self-energy
diagrams in Fig. 1, if we cut along  the internal fermion line. Brackets in the previous expression
represent the thermal  average given by 
\begin{eqnarray}\label{ep11} 
  \langle {\cal{M}}^l_f\,  \rangle 
= \sum_{f} \int{d^{4}p \over (2\pi)^{3}}
 \delta(p^2 - m^2_f)\,{\cal{M}}^l_f\, \eta_{f}(p)
\,.
\end%
{eqnarray}
The previous result shows, that  utilizing  the methods of finite -temperature field theory (at the $g^2$ order),    the neutrino  effective potential reduces to the  
thermal average of neutrino forward scattering amplitude. In fact, it is interesting to combine 
Eqs.  (\ref{ein17}), (\ref{ein19}), and (\ref{ep10})  to write the real part of the index of refraction as 
\begin{eqnarray}\label{ep12} 
 n = 1 +  {2 \pi  \over \kappa^2}  \, \langle {\cal{M}}^l_f \,  \rangle 
\,.
\end%
{eqnarray}
That generalize the zero temperature result \cite{fermi}
 $n = 1 +  2 \pi  \,  {\cal{M}} / \kappa ^2$, by simply  adding the thermal average of
forward scattering amplitude, a result that proves that background does not spoil the coherent condition of the
forward scattering processes.

In the rest of this section we derive  the effective potential  for a 
neutrino propagating in a thermal background composed of electrons, 
protons, neutrons, neutrinos, and their respective anti-particles.
As previously mentioned, the Feynman diagrams  
in Fig. 2  are obtained cutting  the self-energy diagrams of Fig. 1
along the internal fermion line. 
 Diagram (a)  is obtained from the tadpole self-energy, the result is
the same for any neutrino flavor, and one has to sum over all the fermions $f$ present in the
background.   Diagrams  (b) and (c) correspond to cutting the Z- and W-exchange 
self energy diagrams, consequently   the  background fermion necessarily has the 
same flavor as the test neutrino.
  In this way,   ${\cal{M}}^l_f$ can be write as

\begin{equation}\label{ep13}
{\cal{M}}^l_f = {\cal M} _a + {\cal M} _b \delta_{f \nu_l}
+{\cal M} _c \delta_{f l}\,,
\end{equation}
with $ \delta_{ll} =  \delta_{\nu_l \nu_l}  = 1$, $ \delta_{fl} = 0$ 
for $f \neq l$, and $ \delta_{f\nu_l} = 0$ for $f \neq \nu_l$.  Here, 

\begin{eqnarray}\label{ep14}
{\cal M} _a &= & - {1\over 4} \bar u_{\nu_l} (k,s^{\prime}) \Gamma_{f\alpha}^Z u_{\nu_l}(k,s^{\prime}) \Delta^{\alpha \beta}_{Z} (0)
 \bar u_f (p,s) \Gamma_{f\beta}^Z u_f (p,s)\,, \nonumber\\
{\cal M} _b &= & {1\over 4} \bar u_{\nu_l}(k,s^{\prime}) \Gamma_{f\alpha}^Z u_f(p,s)
\Delta^{\alpha \beta}_{Z} (k - p) 
\bar u_f (p,s) \Gamma_{f\beta}^Z u(k,s^{\prime})\,,  \nonumber\\
{\cal M} _c &= & {1\over 4}\bar u_{\nu_l}(k,s^{\prime}) \Gamma_{\alpha}^W u_f(p,s)
\Delta^{\alpha \beta}_{W} (k - p) 
\bar u_f (p,s) \Gamma_{\beta}^W u_{\nu_l}(k,s^{\prime})\,. 
\end{eqnarray}

  Since we are interested in contributions to $V$ 
of order $g^2/M^4_W$  we expand the gauge propagator in power of $M^{-2}_B$
up to the second order
\begin{eqnarray}\label{ep15}
{g_{\alpha\beta}\over q^2 - M_B^2 } \approx - {g_{\alpha\beta}\over M^2_B}(1+ 
{q^2 \over M^2_B } )
\, .
\end{eqnarray}
Using this expansion and neglecting quantities of order   $ m^2_f $/$M_{B}^2 $, we find
\begin{eqnarray}\label{ep16}
{{\cal M}}_a  &\approx &  2 \, \sqrt{2} G_F   \,  X_f k \cdot p \, ,  
\nonumber\\
 {{\cal M}}_b &\approx & \sqrt{2} G_F \, \big[  k \cdot p -
{2 (p \cdot k)^2 \over M_Z^2} \big] \, ,
\nonumber\\
 {{\cal M}}_c &\approx & \sqrt{2} G_F \, \big[  k \cdot p -
{2 (p \cdot k)^2 \over M_W^2} \big]
\, . 
\end{eqnarray}
where $q=p-k$ and  $G_F$/$\sqrt{2} 
\equiv g^2 /8{M_W}^2$ is the Fermi coupling constant, and the 
factors $X_f$ are given in Eqs.   (\ref{ep7}),  (\ref{ep8}), and  (\ref{ep9})   for leptons, neutrinos and nucleons respectively.

Once the previous results are substituted in Eq. (\ref{ep11}), the 
different contributions to $V$ can be expressed in terms of the 
following integrals
\begin{eqnarray}\label{ep17}
\int {d^4 p \over (2 \pi)^3} \, p^\mu  \delta  (p^2 - m^2)
\eta_{f}(p\cdot u)
 &=& A \, u^\mu\, ,  
\nonumber\\
 \int {d^4 p \over (2 \pi)^3} \, p^\mu  p^\nu  \delta  (p^2 - m^2)
\eta_{f}(p\cdot u)
 &= & B \, u^\mu u^\nu\, + \, C \, g^{\mu\nu} 
 \, .
\end{eqnarray}
The scalar quantities   $A, B$, and $C$ are easily evaluated in the rest frame of the medium;
the results are 
\begin{eqnarray}\label{ep18}
A &= &  {1 \over 2} ( N_f -  N_{\bar f}) \,,\nonumber\\
B &=&{1 \over 6} \bigg [ m_f^2 \bigg (  \langle{ 1 \over E_f} \rangle N_f
+  \langle{ 1 \over E_{\bar f} } \rangle N_{\bar f} \bigg) -
   \bigg (  \langle  E_f \rangle N_f
+  \langle  E_{\bar f}  \rangle N_{\bar f} \bigg) \bigg] 
\, ,
\nonumber\\
C& =& - {1 \over 6} \bigg [ m_f^2 \bigg (  \langle{ 1 \over E_f} \rangle N_f
+  \langle{ 1 \over E_{\bar f} } \rangle N_{\bar f} \bigg) -
 4 \bigg (  \langle  E_f \rangle N_f
+  \langle  E_{\bar f}  \rangle N_{\bar f} \bigg) \bigg] 
\, .
\end{eqnarray}
In these  equations  $N_f \, (N_{\bar f})$ represents the density of fermions 
(anti-fermions) in the background
\begin{eqnarray}\label{ep19}
N_{f,\bar f} = g_f \int {d^3p \over (2\pi)^3} {1 \over e^{\beta(E_f 
\mp\mu_{f})} + 1} \, ,
\end{eqnarray}
and $ \langle E_{f,\bar f}^j\rangle$ ( $j=1, -1$) denotes
the statistical averaged of $E_f$ and $1 / E_f$
\begin{eqnarray}\label{ep20}
 \langle E_{f,\bar f}^j\rangle =    {g_f \over N_{f,\bar f}}
\int {d^3p \over (2\pi)^3}  E^j_f {1 \over e^{\beta(E_f
\mp\mu_{f})} + 1} 
\, .
\end{eqnarray}
Here,  $ E_f = \sqrt {p^2 + m_f^2}$ and $g_f$ is the number of spin degrees of freedom ($g_\nu =1$ for 
chiral neutrinos and $g_f= 2$ for the electron and nucleons). Collecting these results 
it straightforward  to write down the effective  potential for a neutrino of a given   flavor.

The  contributions to the effective potential for the 
various neutrino flavors and background particles  are listed  in Table I. They agree with the
results given in the literature \cite{notz,enqv,dnt}.
  
\bigskip\bigskip

{\hspace{2cm} 
\begin{tabular}{|c|c|c|}
\hline  
 $\hbox{Neutrino}$ &  $\hbox {Background Particle}$ & 
$\hbox{$V_{eff}$}$ \\
\hline  $\nu_e, \nu_{\mu}, \nu_{\tau}$  &  $p$ &  
$\pm {G_{F}\over{\sqrt 2}}(1 - 4\sin^2\theta_W) ( N_p - N_{\bar{p}} )$  \\ 
\hline 
  $\nu_e, \nu_{\mu}, \nu_{\tau}$ &  $n$ &  
$\mp{G_{F}\over{\sqrt 2}}(N_n - N_{\bar{n}})$ \\
\hline 
$\nu_e$ & $e$ & $\pm {G_{F}\over{\sqrt 2}}(1 + 4\sin^2\theta_W) 
( N_e - N_{\bar{e}}) \atop +{{2\sqrt{2}G_F{\omega}{m^2_e}}
\over{3M^2_W}}{[N_e\langle {1\over{E_e}}\rangle+N_{\bar{e}}\langle
{1\over{E_{\bar{e}}}}\rangle - {4\over{m^2_e}} (N_e\langle
E_e\rangle+N_{\bar{e}}\langle E_{\bar{e}}\rangle)]}$  \\
\hline
$\nu_e$  & $\nu_e$  &  $\pm {4G_{F}\over{\sqrt 2}}( N_{\nu_e} - 
N_{\bar{\nu}_e} ) - {{8\sqrt{2}G_F{\omega}}\over{3M^2_Z}} [N_{\nu_e}\langle
E_{\nu_e}\rangle+N_{\bar{\nu}_e}\langle E_{\bar{\nu}_e}\rangle] $ \\
\hline
$\nu_{\mu}, \nu_{\tau}$  & $e$  &  $\pm {G_{F}\over{\sqrt 2}}(-1 + 
4\sin^2\theta_W) ( N_e - N_{\bar{e}} )$  \\
\hline
$\nu_{\mu}, \nu_{\tau}$  & $\nu_e$ &  $\pm {2G_{F}\over{\sqrt 2}} 
( N_{\nu_e} - N_{\bar{\nu}_e} )$ \\
\hline
\end{tabular}
}
\vspace{1mm}
{\center\small Table I. Effective potential for a neutrino propagating 
through a medium. The $+$ ($-$) sign refers to neutrinos (anti-neutrinos).
}

\section{Imaginary part}\label{imaginary part}

The discontinuity, of the  self energy is related to the damping 
rate $\gamma$ that determines the imaginary part of the dispersion relation
 (see. Eq. \ref{ein11}), additionally the damping rate can be interpreted
as the rate at which the the single-particle distribution function approaches 
the equilibrium form \cite{wel1}. The former interpretation follows if one consider
a particle distribution that is slightly out of equilibrium, hence one has

\begin{eqnarray}\label{eim1}
{\partial f \over \partial t} = - f \, \Gamma_a  \, +\, \left( 1 + \sigma f \right) \, \Gamma_c
\, ,
\end{eqnarray}
where $\Gamma_a$ and $\Gamma_c$ are the absorption and creation rates of the 
given particle respectively. The parameter $\sigma$ distinguish bosons
($\sigma=1$) and fermions ($\sigma=-1$). The previous equation has for 
general solution 

\begin{eqnarray}\label{eim2}
 f\left( \omega_r, t\right) \, = \, {\Gamma_c \over   \Gamma_a - \sigma  \Gamma_c}
  \, +\, C\left(\omega_r\right) e^{( \Gamma_a - \sigma  \Gamma_c)t} 
\, ,
\end{eqnarray}
where $C\left(\omega_r\right)$ is an arbitrary function that dose not depend on time.
Creation and absorption rates are related by the KMS relation   
\begin{eqnarray}\label{eim3}
  \Gamma_a(\omega) \, = \,   e^{\omega/T} \, \Gamma_c(\omega)
\, .
\end{eqnarray}
Consequently Eq. (\ref{eim2}) can be written as
\begin{eqnarray}\label{eim4}
 f\left( \omega_r, t\right) \, = \, {1  \over   e^{\omega_r/T}- \sigma  }
  \, +\, C\left(\omega_r\right) e^{-2 \gamma t} 
\, ,
\end{eqnarray}
where $\gamma = (\Gamma_a - \sigma  \Gamma_c)/2$ is defined as the damping rate.
Therefore, $\gamma$ can be interpreted as the inverse time scale it takes for a thermal distribution
to reach equilibrium. The sign of the damping rate must necessarily be positive for an 
stable systems.  Additionally, the form of the dispersion relation
implies  for a  normal mode to  propagate that $\gamma$   is small compared to $\omega_r$. 
For neutrinos this conditions is satisfied in normal 
matter, such as the core of the Sun. However, in a CP-symmetric  medium
 the leading contributions to the real part vanish. Thus, 
 the  first non-vanishing  contributions are of order $G_F/M^2_W$; 
under this circumstances $\gamma$ can become of the same order.

To obtain the neutrino damping rate  we have to 
evaluate $\Sigma_{12}$ and then use Eqs. (\ref{ein2}),(\ref{ein5})
and (\ref{ein16}). 
As explained further ahead 
the one loop contributions to  $\Sigma_{12}$ cancel. 
The diagrams that contribute to $\Sigma_{12}$ at the two loops level,
are depicted in Fig. 3; 
there  are also  diagrams similar to those on  
(c) with the $W$ and $Z$ lines interchanged.  According to the 
Feynman rules on  the real time formulation of the FTF, the contributions of diagrams 
(a) and (b) can be written as 
\begin{eqnarray}\label{eim5}
-i{\Sigma}_{12}(k)&=& - \int{{d^4q\over{(2\pi)^4}}} {{d^4p\over{(2\pi)^4}}}
\Gamma^{\mu}_1 (i S_{12}( q)) \Gamma^{\nu}_2  
{\rm Tr}\{\Gamma^{\alpha}_1 (i S_{12}(p))
\nonumber\\
&& \times \Gamma^{\beta}_2 (i S_{21}( p+q-k)) \}
 i (\Delta^{A}_{\mu\alpha}(k-q))_{11}   
i(\Delta^{A}_{\nu\beta} (k-q))_{22}
\, .
\end{eqnarray}
Similarly, for diagrams  (c) and  (d) we have
\begin{eqnarray}\label{eim6}
-i{\Sigma}_{12}(k)&= &\int{{d^4q\over{(2\pi)^4}}} {{d^4p\over{(2\pi)^4}}}
\Gamma^{\mu}_1 (i S_{12}( q)) \Gamma^{\beta}_2 (i S_{21}(q+p-k))  
\Gamma^{\alpha}_1 (i S_{12}(p))\Gamma^{\nu}_2 
\nonumber\\
&&\qquad\qquad\qquad\times  i (\Delta^{A}_{\mu\alpha}(k-q))_{11}  
i (\Delta^{B}_{\nu\beta} (k-p))_{22} 
\, .
\end{eqnarray}
In the above expressions   $A, B=W, Z$ and  $\Gamma^{\alpha}_2= -\Gamma^{\alpha}_1$, with $\Gamma^{\alpha}_1$ representing any of the vertex $\Gamma_{\alpha}^W$ or
$\Gamma^Z_{\alpha f}$ given in the previous section.
The  $S_{12}$ and  $S_{21}$ components of the propagator  matrix 
of the fermion are given by \cite{lands}
\begin{eqnarray}\label{eim7}
S_{12}(p)&=&2\pi i \delta(p^2-m^2)[\eta_f(p)-\theta(-p\cdot u)]
(\splash{p}+m)
\, , 
\nonumber\\
S_{21}(p)&=&2\pi i \delta(p^2-m^2)[\eta_f(p)-\theta(p\cdot u)]
(\splash{p}+m)
\, .
\end{eqnarray}

In principle,   the internal vertices should be added  over the thermal indices ($a=1,2$). 
However,  since  temperature is small as compared to  the gauge boson masses,
 the matrices of the W and Z  bosons   propagators are diagonal  with $(\Delta^{A}_{\alpha\beta} (k))_{11}
=  -  (\Delta^{A}_{\alpha\beta} (k))_{22}^*= \Delta^{A}_{\alpha\beta}$, where 
$\Delta^{A}_{\alpha\beta}$ is the vacuum propagator given in (\ref{ep3}).
This explains the cancellation of the  $g^2$  contribution to the neutrino
damping rate. The one loop contribution to  $\Sigma_{12}$ is given by diagrams  similar
to those in Fig. 1 with  $iS_{12}$ and
 $i(\Delta^{A}_{\alpha\beta} (k))_{12}$ replacing the internal fermionic and
bosonic lines; however the bosonic propagator is diagonal, hence
 $\Sigma_{12}$ cancel at this order. 

According to Eq. (\ref {ein16})  $\gamma$ is directly proportional to  $b_i(\omega, \kappa)$,
that  is given by
\begin{eqnarray}\label{eim8}
b_i(\omega, \kappa) = {1\over4}{\rm Tr}\left\{ \splash{\lambda} \, {\rm Im}\Sigma
\right\}\, ,
\end{eqnarray}
with $ \splash{\lambda}$ defined in Eq.(\ref {ep2}). 

Taking into account the previous results it is demonstrated after a lengthy 
calculation that the neutrino damping rate can be expressed in the form 

\begin{eqnarray}\label{eim9}
\gamma = -i{\epsilon(k\cdot u)\over2in_F} \left\{ 
 C^{WW} +  C^{ZZ} - \sum_{A B=Z,W} D^{AB}\right\}
\, ,
\end{eqnarray}
where $C^{AA}$ is obtained from Eq. (\ref{eim5}):

\begin{eqnarray}\label{eim10}
C^{AA}&=&{1\over 4}\int{{d^4q\over{(2\pi)^4}}} {{d^4p\over{(2\pi)^4}}}
{{d^4Q\over{(2\pi)^4}}}
\bigg\{ {\rm Tr}\{\splash{\lambda}\Gamma^{\mu}_1 (\splash{q}+m_1) 
\Gamma^{\nu}_2\}  {\rm Tr} \{\Gamma^{\alpha}_1 (\splash{p}+m_2)
\nonumber\\
&&\times
\Gamma^{\beta}_2 (\splash{Q}+m_3) \} i (\Delta^{A}_{\mu\alpha}(k-q))_{11}  i (\Delta^{A}_{\nu\beta} (k-q))_{22}
\bigg\}
\nonumber\\
&&\times
(2\pi)^4\delta^{(4)}(Q-(q+p-k)) \delta(q^2-m_1^2) \delta(p^2-m_2^2)\delta(Q^2-m_3^2)
\nonumber\\
&& \times 
[ \eta_F(q_x) - \theta (-q \cdot u)] [ \eta_F(p_y) - \theta(-p\cdot u)]\, [ \eta_F(Q_z) - \theta(Q\cdot u)] 
\, ,
\nonumber\\
\end{eqnarray}
and $D^{AB}$ is obtained from Eq.(\ref{eim6}):
\begin{eqnarray}\label{eim11}
D^{AB} &=&{1\over 4}\int{{d^4q\over{(2\pi)^4}}} {{d^4p\over{(2\pi)^4}}}
{{d^4Q\over{(2\pi)^4}}}
\bigg\{ {\rm Tr} \{\splash{\lambda}\Gamma^{\mu}_1 (\splash{q}+m_1) 
\Gamma^{\nu}_2   (\splash{Q}+m_2)
\nonumber\\
&& \times
\Gamma^{\alpha}_1 (\splash{p}+m_3)\Gamma^{\beta}_2 \}  i (\Delta^{A}_{\mu\alpha}(k-q))_{11}  i (\Delta^{B}_{\nu\beta} (k-p))_{22}
\bigg\}
\nonumber\\
&& \times
 (2\pi)^4\delta^{(4)}(Q-(q+p-k))\delta(q^2-m_1^2) \delta(p^2-m_2^2)\delta(Q^2-m_3^2)
\nonumber\\
&& \times
[\eta_F(q_x)-\theta(-q\cdot u)]  [\eta_F(p_y)-\theta(-p\cdot u)] \, [\eta_F(Q_z)-\theta(Q\cdot u)] 
\,  .
\nonumber\\
\end{eqnarray} 
For later convenience  an extra integration over the momentum $Q$ has been introduced.
In what follows we shall see that the
neutrino damping rate can be expressed in term of amplitudes squared and
weighted with the statistical factors that account for the various physical processes.
To derive these results we notice first  that fermion propagators in 
equations (\ref{eim5}) and (\ref{eim6})  are either type 12 or 21. According
to Eq. (\ref{eim7})  the propagator $S_{12}(p)$ contain a delta function 
$\delta(p^2-m^2)$ and a factor $(\splash{p} + m)$. The delta function    put the fermion  on their mass shell mass,
in other words self-energy diagrams in Fig. 3 are cut along all the internal fermion lines.
Whereas the second factor is concerned,  insertion of the fermion projectors 
 \begin{eqnarray}\label{eim12}
 \splash{p} + m  &=& \sum_s u(p,s) \overline{u}(p,s) 
\nonumber\\
 \splash{p} - m  &=& \sum_s v(p,s) \overline{v}(p,s)  
\, ,
\end{eqnarray}
allow us to rewrite the resulting expressions in terms of amplitudes 
for the physical processes arising from the cuts.  The bosonic
lines are not cut because they do not include thermal distributions
for $T\ll M_W$. 

For definitiveness let us consider  diagram (b) in Fig. 3, and also  that the fermion in the 
internal loop is a proton ($f =   P$). When the diagram is cutted as shown in 
the figure, we obtain a series of physical processes for the  
  neutrinos $\nu_1$, $\nu_2$ and protons $P_1$ and $P_2$. Of these particles 
one of the neutrinos ($\nu^*$) is considered a test particle, all the other are thermalized.
 According to the notation in  Eq. \ref{eim6}, the momentum and chemical potentials are assigned as:
$(\nu_1: k, \mu_{\nu_1})$, $(\nu_2: q, \mu_{\nu_2})$, $(P_1: Q, \mu_{P_1})$,  and$(P_2: p, \mu_{P_2})$.
With momentum and charge conservation  conserved depending of the process, $e.g$ for
 $\nu_1 P_1 \to \nu_2 P_2$: 
 \begin{eqnarray}\label{eim13}
k+Q &= &q + p\, ,
\nonumber\\
 \mu_{\nu_1}+   \mu_{P_1}&=&  \mu_{\nu_1} +  \mu_{P_1}
\, .
\end{eqnarray}
The processes obtained from the mentioned cut rules  include the two neutrinos 
and the two protons distributed into the initial and final states in all possible ways.
Hence, in general we expect to obtain 16 different processes; this is explicitly display in Eq. (\ref{eim16}) 
The resulting expression come out with the appropriated 
thermal distribution, for this we have to rewrite the thermal contributions that 
appear  in Eq. (\ref{eim10}) utilizing the following identities:

\begin{eqnarray}\label{eim14}
\eta_f(x_k)-\theta(\pm{k\cdot u})&=&\mp{\epsilon(k\cdot u )n_f(\mp{
x_k})} \, ,\nonumber\\
n_f(x_k) = e^{-x_k}  n_f(-x_k)& = &e^{-x_k}[1 - n_f(x_k)] \, ,\nonumber\\
{1 \over n_f(x_Q) } n_f(x_k) n_f(x_q)  n_f(x_p) &=& n_f(x_k) n_f(x_q)  n_f(x_p) \nonumber\\
&+& e^{-x_k- x_q+x_p}  \left(1 - n_f(x_k)\right)\left(1 - n_f(x_q)\right) n_f(x_p)
\, , 
\end{eqnarray}
where: 

\begin{eqnarray}\label{eim15}
x_k &=& \beta(k \cdot u - \mu_{\nu_1})\, ,  \qquad x_q = \beta(q \cdot u - \mu_{\nu_2}) \, , \nonumber\\
 x_p &=& \beta(p \cdot u - \mu_{P_1})\, , \qquad  x_Q = \beta(Q \cdot u - \mu_{P_2})
\, . 
\end{eqnarray}
Taking into account these consideration and  performing an integration over the time-like
components of the momentum integration, it is possible to cast the contribution to $\gamma$ arising from   the 
proton loop in diagram (b) of Fig 3 into the following form 
\begin{eqnarray}\label{eim16}
\gamma^{P}&=&{1\over 2\kappa}\int{{d^3q\over{2E_q(2\pi)^3}}} 
{{d^3p\over{2E_p(2\pi)^3}}}
{{d^3Q\over{2E_Q(2\pi)^3}}}(2\pi)^4\bigg[\delta^{(4)}(k + Q - q- p)
\mid{\cal M}(P \nu\leftrightarrow P\nu)\mid^2
\nonumber\\
&&\times [(1-n_{\nu}(E_q))
(1-n_P (E_p))n_P(E_Q)+n_{\nu}(E_q)n_P (E_p)(1-n_P(E_Q))]
\nonumber\\
&&+\delta^{(4)}(k +Q + q - p)
\mid{\cal M}(P\nu\overline{\nu}
\leftrightarrow P)\mid^2 [n_{\bar{\nu}}(E_q) n_P (E_Q)(1-n_P(E_p))
\nonumber\\
&&+(1-n_{\bar{\nu}}(E_q))(1-n_P (E_Q))n_P(E_p)]
+\delta^{(4)}(k + Q+ p -q)
\mid{\cal M}(P\bar{P}\nu \leftrightarrow \nu)\mid^2
\nonumber\\
&&\times [(1-n_{\nu}(E_q))n_{\bar P}(E_p) n_P(E_Q)
+n_{\nu}(E_q)(1-n_{\bar P} (E_p))(1-n_P(E_Q))]
\nonumber\\
&&+\delta^{(4)}(k +Q+p + q)
\mid{\cal M}(P\bar{P}\nu\overline{\nu}
\leftrightarrow 0)\mid^2
[n_{\bar{\nu}}(E_q) n_{\bar P} (E_p)n_P(E_Q)
\nonumber\\
&&+(1-n_{\bar{\nu}}(E_q))(1-n_{\bar P} (E_p)(1-n_P(E_Q))]
+\delta^{(4)}(k -Q-q -p)
\mid{\cal M}(\nu \leftrightarrow P\bar{P}\nu)\mid^2
\nonumber\\
&&\times [(1-n_{\nu}(E_q))(1-n_P(E_p))(1- n_{\bar P}(E_Q))
+n_{\nu}(E_q)n_P (E_p)n_{\bar P}(E_Q)]
\nonumber\\
&&+\delta^{(4)}(k + q-Q-p)
{1\over 2}\mid{\cal M}(\nu\overline{\nu}
\leftrightarrow P\bar{P})\mid^2
[n_{\bar{\nu}}(E_q)(1- n_P (E_p))(1-n_{\bar P}(E_Q))
\nonumber\\
&&+(1-n_{\bar{\nu}}(E_q))n_P (E_p)n_{\bar P}(E_Q)]
+\delta^{(4)}(k + p-Q-q)
\mid{\cal M}({\bar P}\nu
\leftrightarrow \bar{P}\nu)\mid^2
\nonumber\\
&&\times [(1-n_{\nu}(E_q))n_{\bar P}(E_p)(1- n_{\bar P}(E_Q))
+n_{\nu}(E_q)(1-n_{\bar P} (E_p))n_{\bar P}(E_Q)]
\nonumber\\
&&+\delta^{(4)}(k + p + q-Q)
\mid{\cal M}(\bar{P}\nu\overline{\nu}
\leftrightarrow\bar{P})\mid^2
[n_{\bar{\nu}}(E_q)n_{\bar P}(E_p)(1-n_{\bar P}(E_Q))
\nonumber\\
&&+(1-n_{\bar{\nu}}(E_q))(1-n_{\bar P}(E_p))n_{\bar P}(E_Q)]\bigg]
\, .
\end{eqnarray}  
Regardless of its length the interpretation of this equation is quite simple.
The first two terms are interpreted as the absorption and emission of a neutrino via the 
dispersion $\nu P \to \nu P$ with statistical weight 
$ (1-n_{\nu}) (1-n_P )n_P$ and  $ n_{\nu}n_P (1-n_P)$,
respectively.  As expected a $n_f$ factor appears for each background fermion in the initial state, whereas fermion in the final state contribute with a Pauli blocking term $1 - n_f$. As already discussed for femions the absorption  and emission decay rates 
must be added \cite{wel1}. The scattering amplitude  for both processes  are the same, because they are related by CPT inversion. 
Similarly the third and fourth  contributions represents neutrino annihilation  and creation  via the
$P\nu \overline{\nu} \to P$   and $P \to P\nu \overline{\nu} $  respectively; they include, 
as expected,  the statistical factors $ n_{\overline{\nu}}n_P  (1-n_P )$ and  $n_P (1-n_{\overline{\nu}})(1-n_P)  $. 
The same reasoning applies to the remaining terms.

Taking into account the $\delta -$function constrains  some 
of the quoted processes are not  allowed.  In what follows we  focus on 
conditions with temperatures  $m_f \approx T \ll M_W$ where for example  the composition of
the primeval plasma is dominated by (anti-)neutrinos (anti-)electrons,
nucleons and photons.  We recall, see discussion below Eq. \ref{eim4}, that for fermions the contributions
to $\gamma$ of decay and absorption add together. Hence  the statistical factors 
appearing in the previous equation can  be simplified. For example  the absorption $\nu^* P \to \nu P$ and  decay 
$\nu P \to  \nu^* P$, where ($\nu^*$ is the test particle) add according to 

 \begin{eqnarray}\label{eim17}
 (1-n_{\nu}(E_q)) (1-n_P (E_p))n_P(E_Q) &+& n_{\nu}(E_q)n_P (E_p)(1-n_P(E_Q))] 
= \\
& & n_P (E_p)(1-n_P(E_Q))] + n_{\nu}(E_q) \left(  n_P (E_p)- n_P (E_Q) \right)
\, . \nonumber
\end{eqnarray}
However we can drop out the last term in the right hand side of the equation, because
its contribution vanish when substituted into Eq. (\ref{eim16}).

With  all these results we finally find that the neutrino damping rate 
can be expressed as 
\begin{eqnarray}\label{eim18}
\gamma&=&\pm {(2\pi)^4\over{2\omega}} \bigg\{{1\over 2}
\int{{d^3p\over{(2\pi)^3}2E_p} {{d^3q\over{(2\pi)^32E_q}}}{{d^3Q
\over{(2\pi)^32E_Q}}} }\mid  {\cal{M}}(\nu 
e \leftrightarrow\nu e) \mid^2 \nonumber\\ && \qquad\qquad \qquad
\times \delta^2(k+p-q-Q)(1-n_{e}(E_Q))n_{e}(E_p) \nonumber\\ && \qquad
\qquad \qquad +\, \hbox{\rm { all the possible processes}}\bigg\} \,
,
\end{eqnarray}
where the $+$ ($-$)  sign stands for test neutrinos
(anti-neutrinos). For all possible 
processes we mean all the kinematically allowed processes  obtained by cutting all the
fermionic internal lines of Feynman diagrams shown in Fig. 3. 
The corresponding processes and their cross sections are listed below in 
(\ref{eim22}).

The damping rate can be written in terms of the thermal average 
of the  cross section
For the dispersion $\nu f\leftrightarrow \nu f$, the  differential
cross section is  given by 
\begin{eqnarray}\label{eim19}
 d \sigma_f={1\over V_{\rm rel}2\omega 2E_p} \mid
{\cal{M}}(\nu f \to\nu f)\mid^2(2\pi)^4 \delta^2(k+p-q-Q) 
{{d^3q\over{(2\pi)^32E_q}}}{{d^3Q\over{(2\pi)^32E_Q}}} \, ,
\end{eqnarray}
where $ V_{\rm rel}$ is  the relative velocity between the  neutrino and the 
background fermion; as we are considering massless neutrinos we simply have  
$ V_{\rm rel}= 1$.

Thus, $\gamma$ reads
\begin{eqnarray}\label{eim20}
\gamma = \pm  \sum_f a_f \int{d^3 p\over{(2\pi)^3}} \langle d\sigma_f\rangle 
\, n_f(E_p) \, , + ..........
\end{eqnarray}
where $a_f=1/2$ for $f=e,n,p$  and  $a_f=1/4$ for $f=\nu$  and 
\begin{eqnarray}\label{eim21}
\langle d\sigma_f\rangle ={(2\pi)^4\over{2\omega 2E_p v_{\rm rel}}}
\int{{d^3q
\over{(2\pi)^32E_q}}} {{d^3Q\over{(2\pi)^32E_Q}}}\,\, \mid
{\cal{M}}\mid^2\,\, \delta^2(k+p-q-Q) 
(1-n_f(E_Q)) \, .
\end{eqnarray}
is the cross section thermally  averaged by the Pauli blocking term   ($1-n_{f}(E_Q)$). 
In Eq. (\ref{eim18}) the ellipsis represent the other possible process, in each case the 
corresponding statistical factors in Eqs. (\ref{eim18}, \ref{eim19}) and the dispersion amplitude
are replaced by the pertinent factors.
We recall that according to Eq. (\ref{ein19}) $\gamma$ is directly related to the imaginary 
part of the index of refraction. Hence, Eq. (\ref{eim20}) can be identified with the 
optical theorem.

In what follows we shall  apply the previously obtained results to explicitly 
 evaluate the neutrino damping rate in a background composed of neutrinos,
electrons, protons and neutrons.  We suppose that  the Pauli  blocking  term
can be neglected,  in  addition we consider temperatures $m_f \approx T \ll M_W$, hence 
in the thermal averages we can assume that $q^2 \ll M^2_W$. 
 First let us consider the 
cross section for the relevant  processes. It is  common to quote the cross sections, assuming 
ultrarelativistics neutrinos and   neglecting the fermion masses. However for conditions as those
 of the  early universe, temperature and consequently the average neutrino energy can be comparable to the nucleon masses, and sometimes to the lepton masses.  Hence, keeping  fermion masses, the various neutrino cross sections
can be calculated as 
\begin{eqnarray}\label{eim22}
\tilde{\sigma}(\nu_e e\leftrightarrow\nu_e e)&=&16\delta^2+ 12 \delta 
+3 -(40 \delta^2+30\delta+6){m_e^2\over s}+ 
(12 \delta^2+8\delta+1){3m_e^4\over s^2}
\nonumber\\
&-&(16 \delta^2+2\delta){m_e^6\over s^3}+ 4\delta^2 {m_e^8\over s^4}\, ,
\nonumber\\
\tilde{\sigma}(\nu_{\mu,\tau} e\leftrightarrow\nu_{\mu,\tau} e)&=&16{\delta}^2
- 12 \delta +3 -(26 \delta^2-15\delta+3){2m_e^2\over s}+ 
(12 \delta^2-8\delta+1){3m_e^4\over s^2}
\nonumber\\
&-&(8 \delta^2-3\delta){2m_e^6\over s^3}+ 4\delta^2 {m_e^8\over s^4}\, ,
\nonumber\\
\tilde{\sigma}(\nu_e \bar{e}\leftrightarrow\nu_e \bar{e})&=&16{\delta}^2+
4\delta +1 -(40 \delta^2+10\delta+1){m_e^2\over s}+ 
(12 \delta^2-4\delta){3m_e^4\over s^2}
\nonumber\\
&-&(16 \delta^2+10\delta+1){m_e^6\over s^3}+ (2\delta+1)^2 {m_e^8\over s^4}\, ,
\nonumber\\
\tilde{\sigma}(\nu_{\mu,\tau} \bar{e}\leftrightarrow\nu_{\mu,\tau} \bar{e})
&=&16{\delta}^2- 4\delta +1 -(40 \delta^2-16\delta+1){m_e^2\over s}+ 
(12 \delta^2-4\delta){3m_e^4\over s^2}
\nonumber\\
&-&(16 \delta^2-12\delta+1){m_e^6\over s^3}+ (-2\delta+1)^2 {m_e^8\over s^4}\, ,
\nonumber\\
\tilde{\sigma}(\nu_i \nu_i\leftrightarrow\nu_i \nu_i)&=& 12\, ,
\nonumber\\
\tilde{\sigma}(\nu_i\bar{\nu}_i\leftrightarrow\nu_i \bar{\nu}_i)&=&8\, ,
\nonumber\\
\tilde{\sigma}(\nu_i \nu_j\leftrightarrow\nu_i \nu_j)&=&6 \, ,
\nonumber\\
\tilde{\sigma}(\nu_i \bar{\nu}_j\leftrightarrow\nu_i \bar{\nu}_j)&=&2\, ,
\nonumber\\
\tilde{\sigma}(\nu_i n\leftrightarrow\nu_i n)&=&3-{6m^2_n \over s} 
+ {3m^4_n\over s^2} \, ,
\nonumber\\
\tilde{\sigma}(\nu_i \bar{n}\leftrightarrow\nu_i \bar{n})&=&2-{2m^2_n \over s} 
- {2m^6_n\over s^3}  \, ,
\nonumber\\
\tilde{\sigma}(\nu_i p\leftrightarrow\nu_i p)&=&(16{\delta}^2-12 \delta 
+1)-(40{\delta}^2- 30 \delta +6){m^2_p\over s}+(36{\delta}^2
- 24 \delta +1){m^4_p\over s^2}
\nonumber\\
&-&(16{\delta}^2- 6 \delta){m^6_p\over s^3}+(4{\delta}^2- 
2\delta){m^8_p\over s^4}  \, ,
\nonumber\\
\tilde{\sigma}(\nu_i \bar{p}\leftrightarrow\nu_i\bar{p})&=&(16{\delta}^2-4 
\delta +1)-(40{\delta}^2- 10 \delta +1){m^2_p\over s}+(36{\delta}^2
-12 \delta){m^4_p\over s^2}
\nonumber\\
&-&(16{\delta}^2- 10\delta+1){m^6_p\over s^3}+(4{\delta}^2- 
2\delta){m^8_p\over s^4}  \, ,
\nonumber\\
\tilde{\sigma}(\nu_e \bar{\nu}_e \leftrightarrow e \bar{e})
&=&\left( 1-{4m^2_e\over s} \right)^{1/2} \left( 16{\delta}^2+ 8 \delta +2
+ (8\delta^2 + 4\delta -{1\over 2}){4m^2_e\over s} \right) \, ,
\nonumber\\
\tilde{\sigma}(\nu_{\mu,\tau} \bar{\nu}_{\mu,\tau} \leftrightarrow e \bar{e})
&=&\left(1-{4m^2_e\over s}\right)^{1/2}\left(16{\delta}^4- 8 \delta +2
+ (16\delta^2 -8\delta +{3\over 2}){2m^2_e\over s}\right)\, , 
\end{eqnarray}
where $\tilde{\sigma}=\sigma/\sigma_0$ with $\sigma_0={G_F^2\over 12\pi}s$, 
$i=e,\mu,\tau$,  $s=(k+p)^2$ is the Mandelstam variable, and
$\delta=\sin^2\theta_W  \approx 0.229$. These results reduce, in the zero fermion limit,
to those found in Enqvist, Kainulainen and Thomson \cite{enqv2} and Langacker and Liu \cite{lang2}

Once  the cross sections are inserted in Eq. (\ref{eim20}) the thermal averages 
should be evaluated according to the constrains of the problem.  If  
we consider temperatures well bellow the nucleon mass,  then the
proton and neutron  contributions will be suppressed by the Boltzmann factor, and their 
contribution  neglected.
On the other hand if $T \approx m_f$ the complete average of the cross sections in Eq. (\ref{eim22}) should be
considered. In what follows we  consider situation in which $m_f < T$ and we retain terms of order 
$m_f^2/ s$ in (\ref{eim22}). This 
 lead us  to consider thermal
averages that contain integrals of the following type
\begin{eqnarray}\label{eim23}
\int {d^4 p \over (2 \pi)^3} \, \left( k \cdot p\right)\left( k \cdot u\right)
\theta( k \cdot u ) \delta  (p^2 - m_f^2)
 n_f(p)
 \, .
\end{eqnarray}
These integrals are easily evaluated in terms of the thermal average of the 
fermion density and  energy, utilizing the Eqs. (\ref{ep17}-\ref{ep20}).
The results can be collected in a general formula that gives the contributions to the 
neutrino damping rate arising form various background particles (to first order in $m_f$):

\begin{eqnarray}\label{eim24}
\gamma_{\nu_i} =  {G_F^2 \over 12 \pi} \sum_{f} \, \,  {a_f \, N_f\over g_f}
\left\{ 2\omega A_f  \langle{ E_f }\rangle 
+ m_f^2 \left( A_f + B_f \right)  \right\}
\, ,
\end{eqnarray}
where $\omega$ is the neutrino energy,  summation in $f$   is taken over fermions in the background, and the corresponding  factors $A_f$ and $B_f$ are summarize 
in Table 2. In this equation $N_f$ represents the density of fermion of antifermions in the medium (Eq. \ref{ep19})
and the statistical averages of $E_f$  is defined in Eq. (\ref{ep20}). 

\vspace{1.5cm}
\begin{tabular}{|c|c|c|c|} 
\hline
$\hbox{ neutrino}$ & $\hbox{Background fermion $f$}$ & 
$\hbox{$A_f$}$ & $\hbox{$B_f$}$ \\
\hline
$\nu_e$ & $e$  & 16 $\delta^2 + 12\delta + 3$  &  
$-(40\delta^2 +30\delta +6)$  \\
\hline
$\nu_{\mu,\tau}$ & $e$  & 16 $\delta^2 - 12\delta + 3$  &  
$-(52\delta^2 -30\delta +6)$  \\
\hline
$\nu_e$ & $e^{-}$ &  $16 \delta^2 + 4\delta + 1$ &  
$-(40\delta^2+10\delta+1)$ \\
\hline
$\nu_{\mu,\tau}$ & $e^{-}$ &  $16 \delta^2 - 4\delta + 1$ &  
$-(40\delta^2-16\delta+1)$ \\
\hline
$\nu_i$ & $\nu_i$ &  $12$ &  $0$  \\
\hline
$\nu_i$ & $\nu_j$, $i\neq j$ &  $6$ &  $0$  \\
\hline
$\nu_i$ & $\overline{\nu}_i$  &  $8$  &  $0 $ \\
\hline
$\nu_i$ & $\overline{\nu}_j$, $i\neq j$  &  $2$  &  $0 $ \\
\hline
$\nu_i$ & $n$  &  $3$  &  $-6$  \\
\hline
$\nu_i$ & $\overline{n}$  &  $2$  &  $-2$  \\
\hline
$\nu_i$ & $p$  &  $16\delta^2 - 12\delta + 1$   &  
$-(40\delta^2 - 30\delta + 6$ \\
\hline
$\nu_i$ & $\overline{p}$ & $16\delta^2 - 4\delta + 1$  &  
$-(40\delta^2-10\delta+1)$\\
\hline
$\nu_e \overline{\nu}_e  \leftrightarrow  e \overline{e}$ & 
\hbox{annihilation  process} &
$16\delta^2 + 8\delta + 2$  & $- 6$ \\
\hline
$\nu_{\mu,\tau} \overline{\nu}_{\mu,\tau} \leftrightarrow e\overline{e} $& 
\hbox{annihilation  process} &
$16\delta^2 - 8\delta + 2$  & $-1$ \\ 
\hline
\end{tabular}

\vspace{1mm}
{\center\small Table II. Coefficient $A_f$ y $B_f$ appearing in Eq. (\ref{eim24}) for various  processes between neutrinos and the quoted background particles. Here  
$\delta = {\rm sen}^2\theta_W $.}
\vspace{1.5cm}

The next step is to  quote some explicit results. Consider a CP-symmetric plasma composed 
the three type of neutrinos $\nu_e$, $\nu_\mu$ and $\nu_\tau$, electrons and their 
corresponding antiparticles.  Considering a $\nu_e$ as  a test particle, the contribution to  the neutrino damping rate arising from the background neutrinos  is given by  
\begin{eqnarray}\label{eim25}
\gamma = 8.1   {G_F^2 \over \pi^3} \, \omega \,  T^4 \, , 
\end{eqnarray}
where $\omega$ is the neutrino energy. Whereas  the contribution of the 
electron and positron background  to the $\nu_e$ damping rate yields
\begin{eqnarray}\label{eim26}
\gamma =  0.39 {G_F^2 \over  \pi^3} \, \omega \,  T^4  \rho \left( {m_e \over T}  \right) 
\left[  \epsilon \left( {m_e \over T} \right)  
- 0.6  {m_e^2 \over \omega T}     \right] , 
\end{eqnarray}

where the functions $\rho$ and $\epsilon$ are defined by

\begin{eqnarray}\label{eim27}
\rho (\xi) &=&  \int_{0}^{\infty}   dx  {x^2 \over   e^{\sqrt {x^2 + \xi^2 } }  + 1 }   \, , \nonumber \\
   \epsilon (\xi) &=& {1 \over \rho (\xi) }  \int_{0}^{\infty}   dx  {x^3 \over   e^{\sqrt {x^2 + \xi^2 } }  + 1 }
\, . 
\end{eqnarray}
Similar expressions can be obtained for the other processes. These results reduce 
to those in references \cite{notz,enqv2,lang2} in the limit of zero electron mass.

\section{conclusions}

In this paper we have systematically derived  the neutrino damping rate in real-time
thermal field theory.  Starting from the discontinuity of the neutrino self 
energy at the two loop level, we prove that the damping rate can be expressed as integrals
over space phase of total cross sections, weighted with statistical factors that 
account for the possibility of particle absorption or emission from the medium.

Cutkosky rules are used  to obtain the self-energy  imaginary part at zero temperature. 
 Weldon \cite{wel1} and  Kobes and Semenoff  \cite{kob1} (see also \cite{jizba}) have studied 
the corresponding Cutkosky rules
 at finite temperature. In these references it is 
shown that  for  certain specific examples at one loop order the 
discontinuity of the self-energy  can be expressed in term of amplitudes squared and
weighted with the statistical factors that account for the various physical processes.
Here we  prove that these result stand valid for the neutrino damping rate
at the two loop order.

The complete  results that account for all possible processes  that contribute to $\gamma$ appear 
in Eq. (\ref{eim16}). Depending of the physical conditions some of these processes
are forbidden.  Specific results for conditions such as those of the early universe where 
the primeval plasma is composed of (anti-)neutrinos, (anti-)electrons, and nucleons 
were obtained.  For those conditions the fermion masses are not always negligible,
consequently we report a general formula (Eq. \ref{eim24}) that includes 
mass correction to first order in $m_f^2/  \langle s \rangle $; however,   further improvements are easily  obtained
utilizing the values for the cross sections
in Eq. (\ref{eim22}). Our results, 
summarized in Eq. (\ref{eim24}) and Table II, should be useful for  the 
study of neutrino processes in the early universe, as well in some astrophysical 
scenarios.

\section*{acknowledgment}
This work has been partially supported by CIC-UMSNH, grant 8.10.
and by CONACyT grants No. 32540-E and G32723-E.

\begin{figure}[h]
\vspace{4mm} \epsfxsize=10.5cm \centerline{\epsffile{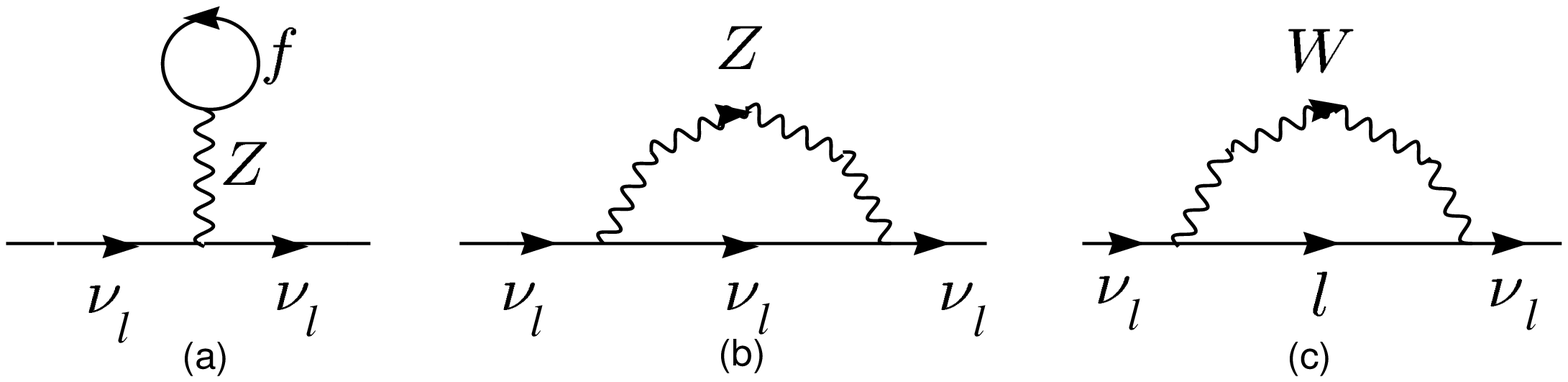}} 
\vspace{3mm} 
\caption{One-loop diagrams for the self-energy of a neutrino
in a thermal background of charged leptons nucleons and neutrinos. In (a), $f$ stands for any  fermion in the background. In (c), the charged  lepton $l$ is of the same flavor as the neutrino.}
\end{figure}

\begin{figure}[h]
\vspace{4mm} \epsfxsize=10.5cm \centerline{\epsffile{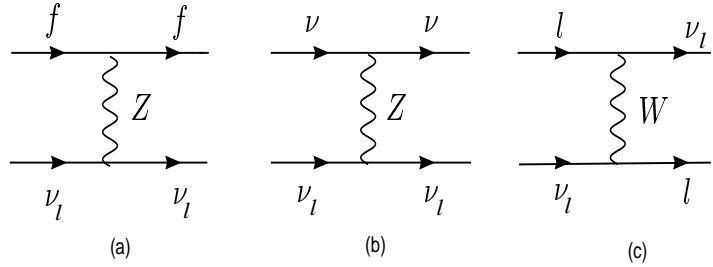}} 
\vspace{3mm} 
\caption{Tree-level diagrams for $\nu_lf \to \nu_lf$ forward  scattering; $f$ represents the  background fermions. These diagrams are obtained when the corresponding self-energy diagrams [Fig. 1] are cutted along the internal fermion line.}
\end{figure}

\begin{figure}[h]
\vspace{4mm} \epsfxsize=10.5cm \centerline{\epsffile{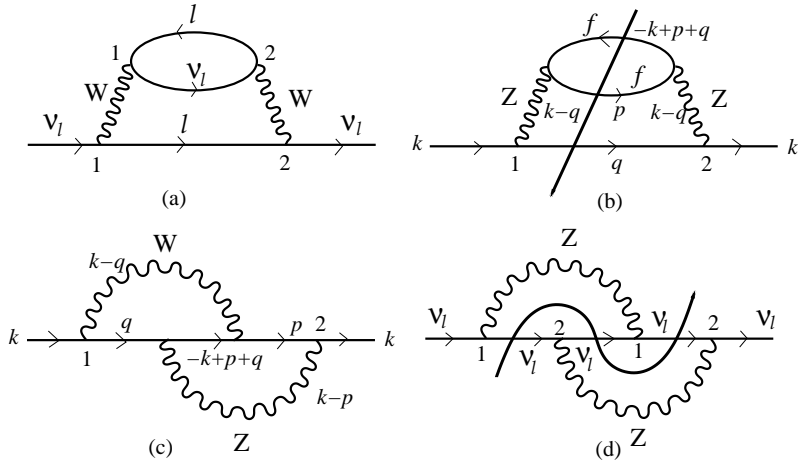}} 
\vspace{3mm} 
\caption{ Two-loops contributions to  the self-energy of a neutrino in 
a thermal background of charged leptons, nucleons, and neutrinos.}
\end{figure}

\end{document}